\documentclass[12pt]{article}
\usepackage[dvips]{graphicx,psfrag}
\usepackage{amssymb}
\input epsf
\newcommand{\be}{\begin{equation}}
\newcommand{\ee}{\end{equation}}
\newcommand{\bea}{\begin{eqnarray}}
\newcommand{\eea}{\end{eqnarray}}
\newcommand{\lb}{\label}
\begin{document}
\begin{titlepage}
\begin{center}
{\large\bf  CAN PRIMORDIAL BLACK HOLES BE A SIGNIFICANT PART OF DARK MATTER?}
\vskip 1cm
{\bf David Blais}
\vskip 0.4cm
 Laboratoire de Physique Math\'ematique et Th\'eorique,
 UMR 5825 CNRS,\\
 Universit\'e de Montpellier II, 34095 Montpellier, France.\\
 E-mail: {\tt blais@LPM.univ-montp2.fr}
\vskip 0.7cm
{\bf Claus Kiefer}
\vskip 0.4cm
 Institut f\"ur Theoretische Physik, Universit\"at zu K\"oln,\\
 Z\"ulpicher Str.~77, 50937 K\"oln, Germany.\\
 E-mail: {\tt kiefer@thp.uni-koeln.de}
\vskip 0.7cm
{\bf David Polarski}
\vskip 0.4cm
 Laboratoire de Physique Math\'ematique et Th\'eorique,
 UMR 5825 CNRS,\\
 Universit\'e de Montpellier II, 34095 Montpellier, France.\\
\vskip 0.3cm
 Laboratoire de Math\'ematiques et Physique Th\'eorique,
 UMR 6083 CNRS,\\
  Universit\'e de Tours, Parc de Grandmont, 37200 Tours, France.\\
 E-mail: {\tt polarski@LPM.univ-montp2.fr}

\end{center}
\date{\today}
\vskip 2cm
\begin{center}
{\bf Abstract}
\end{center}

\begin{quote}
The computation of PBH (primordial black hole) production 
from primordial perturbations has recently been improved by considering
a more accurate relation between the primordial power spectrum
and the PBH mass variance. 
We present here exact expressions which are valid for 
primordial spectra of arbitrary shape and which allow accurate 
numerical calculations. 
We then consider the 
possibility to have a significant part of dark matter in the form of PBHs 
produced by a primordial spectrum of inflationary origin possessing a 
characteristic scale. We show that in this model the relevant PBH mass 
is constrained to lie in the range 
$5\times 10^{15}\ {\rm g}\lesssim M \lesssim 10^{21}\ {\rm g}$. 
This is much less than the mass range coming from the QCD phase transition, 
allowing the two mechanisms to be easily distinguished.
\end{quote}

\end{titlepage}

\noindent
{\it Introduction}:
A consistent paradigm seems to emerge in cosmology, of which dark matter is an 
essential ingredient. However, the nature of dark matter remains to be
one of the most important open problems. It is expected to 
account for about one quarter of the present critical energy density,
most of the remaining three quarters being dark energy whose nature
is also unknown. There exist many
candidates which are classified into cold, hot, and warm dark matter. 
Some of these candidates would signal new physics beyond 
the standard model of particle physics,
such as relic neutralinos, the supersymmetric cold dark matter
(CDM) candidate.
Another possible CDM candidate could be primordial black holes (PBHs). 
The advantage with them is that their existence rests on known physics,
general relativity and the presence of primordial fluctuations, independent
of the mechanism which generates them. 
Indeed, it is well known that PBHs can be produced in the early 
universe due to the collapse of overdense regions on the linear size of the 
Hubble radius at the time when this corresponds to their Schwarzschild radius
\cite{CH74}. 
These fluctuations in the energy density must exist one way or the other in 
order to explain the origin of all inhomogeneities we see in the universe. 
Their existence, and in particular their primordial origin, is reflected in 
the presence of acoustic peaks in the cosmological microwave background
(CMB) anisotropy, and is now well established by observations. 
Therefore, the generation of PBHs is unavoidable.
Inflationary fluctuations are produced on a huge range of scales
and the observation of PBHs 
formed after inflation could probe the fluctuations on scales
where the CMB gives no information.
It would therefore be complementary to the results which are extracted 
from the CMB data.

PBHs were produced in the very early universe. PBHs generated 
before $10^{-23}\ {\rm s}$, corresponding to masses
$M<M_*\approx 5\times 10^{14}\ {\rm g}$, have already evaporated by the 
present day due to Hawking radiation. PBHs with masses
bigger than $M_*$ could, however, constitute 
a significant fraction, or even all, of the CDM.
 Unfortunately this 
possibility, though attractive, cannot be implemented with
 scale-free perturbations because they would lead to a negligible
rate of PBH formation. 
However, in view of the naturalness of PBH generation,
it is important to 
investigate whether other kinds of spectra could lead to a significant
formation rate. 
 
\vskip 10pt
\par\noindent
{\it General formalism}: We will consider Gaussian primordial fluctuations, 
the kind of fluctuations expected in most inflationary models. The density 
contrast averaged over a sphere of radius $R$ then reads
\be
p(\delta) = \frac{1}{\sqrt{2\pi}~\sigma (R)}~ e^{-\frac{\delta^2}
{2 \sigma^2(R)}}\ ,
\ee 
where the dispersion $\sigma^2(R) \equiv \Bigl \langle \Bigl ( \frac{\delta M}{M}  
\Bigr )_R^2 \Bigr \rangle$ is computed using a top-hat window function, 
\be
\sigma^2(R) = \frac{1}{2\pi^2}\int_0^{\infty}dk ~k^2
 ~W^2_{TH}(kR) ~P(k)\lb{sigW}~.
\ee
$P(k)$ is the primordial power spectrum, and $W_{TH}(kR)$
is the Fourier transform 
of the top-hat window function divided by the probed volume
 $V_W=\frac{4}{3}\pi R^3$,
\be
  W_{TH}(kR)=\frac{3}{(kR)^3}\bigl (\sin kR-kR\cos kR\bigr )\ .
\ee
The probability $\beta(M_H)$ that a region
 of comoving size $R=\frac{H^{-1}}{a}|_{t_k}$ has an averaged 
density contrast at horizon crossing $t_k$ in the range 
$\delta_{min}\leq\delta\leq\delta_{max}$, is given by
\be
\label{beta}
  \beta(M_H)=\frac{1}{\sqrt{2\pi}\,\sigma_H(t_k)}\,
           \int_{\delta_{min}}^{\delta_{max}}\,
           e^{-\frac{\delta^2}{2 \sigma_H^2(t_k)}}\,\textrm{d}\delta
          \approx\frac{\sigma_H(t_k)}{\sqrt{2\pi}\,\delta_{min}}
           e^{-\frac{\delta_{min}^2}{2 \sigma_H^2(t_k)}}\ ,
\ee
where $\sigma_H^2(t_k)\equiv \sigma^2(R)\big|_{t_k}$. 
Recently, the connection between the mass variance on very small scales
and the normalization 
on present COBE scales has been improved yielding a substantially
lower mass variance \cite{BKP1}. 
This formalism also allows for an accurate calculation
of the mass variance for spectra with features.
The quantity 
$\delta^2_H(t_k)\equiv \frac{k^3}{2\pi^2}~P(k,t_k)=
\frac{2}{9 \pi^2}~k^3~\Phi^2(k,t_k)$ 
is often used; it is related to $\sigma_H^2(t_k)$
for an arbitrary primordial power spectrum as \cite{P02,BBKP}
\be
\sigma_H^2(t_{k})=\delta^2_H(t_k)~\int_0^{\infty}
 x^{3}~\frac{F(kx)}{F(k)}~T^2(kx,t_k)~
                                            W^2_{TH}(x)~\textrm{d}x~,\lb{sigF1}
\ee
where $k^3 \Phi^2(k,t_k)\equiv \frac{4}{9}~F(k),~t_k\ll t_{eq}$
(radiation-dominated phase).
The time-independent quantity $F(k)$ 
defines the primordial spectrum on super-Hubble radius scales. 
The upper integration limit can safely be taken to be infinity
due to the rapid convergence of the integral.
The evolution of the fluctuations 
on small scales differs from the long wave evolution,
and this difference is encoded in the transfer function 
$T(k',t)$. At the present day we have $t_k=t_0,~k=k_0=a_0 H_0$ and
 $T(k',t_k)\neq T(k',t_0)$. This is why the 
relation between $\sigma_H^2(t_k)$ and $\delta^2_H(t_k)$
 is {\it not} the same as today. 
An accurate evaluation of the transfer function
will be presented elsewhere \cite{BBKP} and gives
\be
T^2(kx,t_k)=W_{TH}^2(c_s x)~,~~~~~~~~~~~~~~~~c_s^2=\frac{1}{3}\lb{T}~.
\ee
A pure scale-invariant spectrum corresponds to
 $F(k)={\rm constant}$, and a numerical estimate for a critical-density 
universe ($\Omega_{m,0}=1$) gives \cite{BBKP}
\be
\sigma_H^2(t_{k}) = 5.37~\delta^2_H(t_k) = 6.63~\delta^2_H(t_0)~.\lb{n1}
\ee
The prefactors in the second and last equality differ
 by a factor $\left (\frac{10}{9} \right )^2$ 
which takes into account the transition from radiation to matter domination. 
In particular, the constraint on a pure blue spectrum
(spectral index $n>1$) is weakened.
The inclusion of a cosmological constant $\Lambda$
 with $\Omega_{\Lambda,0}=0.7$ reduces the mass variance even further
 by about 15\% \cite{P02}.
This constant decrease is degenerate with an increase in $\delta_{min}$
\cite{BBKP}, so we choose to make 
our numerical simulations for the fiducial case
 $\Omega_{m,0}=1,~\delta_{min}=0.7$.
However, there is another important aspect of (\ref{sigF1}).
 For a spectrum with a feature at 
some scale $k_s$, the shape of the power spectrum will translate
 in a non-trivial way into 
$\sigma_H^2(t_k)$ (see, e.g., Fig.~1 below). 
Though the refinements described here make the prospects
 for the cosmological relevance of PBHs 
less promising since it reduces the probability to form PBHs,
the situation looks bad only for scale-free spectra.
 Indeed, potentials with a feature on small scales could dramatically 
improve the PBH formation rate.
 This possibility was already considered in the past \cite{INN94,RSG96} 
and we would like to take it up again in the light of the
improved calculations done in \cite{BKP1}.

\vskip 10pt
\par\noindent
{\it PBH as a CDM candidate}:
 It is not possible to have a significant fraction of the dark matter 
in the form of PBHs with a scale-free powerlaw primordial spectrum. 
We assume that the probability $\beta(M)$
 represents the fraction of the energy density 
at the PBH formation time that is going to form PBHs with mass $M$.
Their present contribution to the critical density, $\Omega_{PBH,0}$, 
is then given by \cite{BKP1}
\be
\Omega_{PBH,0}(M)~h^2 = 6.35 \times 10^{16}~\beta(M) 
           ~\left(\frac{10^{15}\textrm{g}}{M}\right)^\frac{1}{2}~.\lb{omega}
\ee
For a pure Harrison-Zel'dovich spectrum normalized to COBE,
 the mass variance at 
Hubble-radius crossing, and therefore $\beta(M)$,
 is (nearly) constant on all mass scales. 
When we go to lower masses, the 
corresponding $\Omega_{PBH,0}$ increases $\propto M^{-\frac{1}{2}}$
 because PBHs of 
lower mass form earlier in the radiation 
era, while their energy density decreases less rapidly than that of radiation. 
On the other hand, the COBE normalization of the spectrum results in 
$\beta(M)\ll 10^{-17}$ so 
that this increase is by far insufficient and will result in a negligible 
contribution of PBHs to the present energy density. 
One way to circumvent the problem is to invoke 
a blue spectrum, $n>1$, in which case $n\approx 1.3$ is needed in order 
to have a significant contribution with PBHs of mass $M_*$ \cite{GL,BKP1}. 
In such models, the spectral index is assumed to be constant
over a huge range of scales.  
A more serious problem is that as the Hubble mass
at the end of inflation is lower 
than $M_*$, a sharp constraint exists on the density of evaporated PBHs
from their contribution 
to the $\gamma$-ray background, $\Omega_{PBH,0}\sim 10^{-8}$.
Therefore, a blue spectrum that satisfies the bound from the 
$\gamma$-ray background necessarily yields $\Omega_{PBH,0}\ll 1$. 
The only way out is to boost the production of PBHs in a localized way. 

Microlensing gives no detection limits on the mass of massive 
compact halo objects (MACHOs)
 below $10^{-7}\ {\rm M}_{\odot}$. On the other hand, 
it is advantageous to form PBHs earlier, with masses not too far from $M_*$. 
For larger masses, larger values for $\beta$ are required
 in order to achieve the same 
$\Omega_{PBH,0}$. We are thus looking for a spectrum
 which increases $\beta(M)$ in this region, 
but {\it not} in the mass range $M_{PBH}\leq M_*$.
 This cannot be achieved neither with a blue 
spectrum, nor using a pure step-like spectrum with more power on small scales
\cite{BKP1}. 
In the following we shall present a model
with a feature at some characteristic scale, where this is possible.

\vskip 10pt
\par\noindent
{\it The model}: We consider here a model based on an
inflaton potential with a step in its first derivative. 
The adiabatic perturbations spectrum was derived analytically
by Starobinsky \cite{S92} and 
has been applied either to explain a possible bump
in the matter power spectrum $P(k)$ \cite{LPS98} on a scale 
$k\approx 125 h^{-1}$Mpc or on small scales as a cure
to the dwarf galaxy problem \cite{KL00}. 
In our case the feature will be placed on even much smaller scales,
typically the scale corresponding to PBHs with mass $M_*$. 
We expect that a deeper understanding of fundamental theories
could give a distinguished scale from which the PBH formation
rate could be calculated.
In this model, the spectrum is defined by a characteristic scale
 $k_s$ (or mass $M_s$) and a parameter  
$p$, where $p^2$ is the ratio of the power spectrum on large scales with respect to that on small scales. 
In view of its rich structure in the vicinity of $M_s$, 
in particular its oscillations, one must use (\ref{sigF1}) 
in this crucial part of the spectrum in order to compute
$\sigma_H(M)$ correctly. 
The probability $\beta(M_H)$
 to produce a PBH with mass $M$ is obtained from the 
primordial spectrum in a non-trivial way according to
\be\lb{sig}
\sigma_H^2(M)=\frac{8}{81\pi^2} \int_0^{\infty} x^3~F(b x)~
                                    W^2_{TH}(c_sx)~W^2_{TH}(x)~\textrm{d}x~,
~~~~~~~~b\equiv \frac{\sqrt{M_s}}{\sqrt{M}}
\ee   
with $M\equiv M_H(t_k), M_H(t_{k_s})\equiv M_s$, and $t_k$
 is the Hubble radius (``horizon'') crossing time $k= (aH)_{t_k}$. 
The non-trivial way in which the primordial spectrum $F(k)$
 translates into the mass 
variance $\sigma_H(M)$ is obvious from Fig.~1.

\begin{figure}
  \begin{center} 
\mbox{ \input{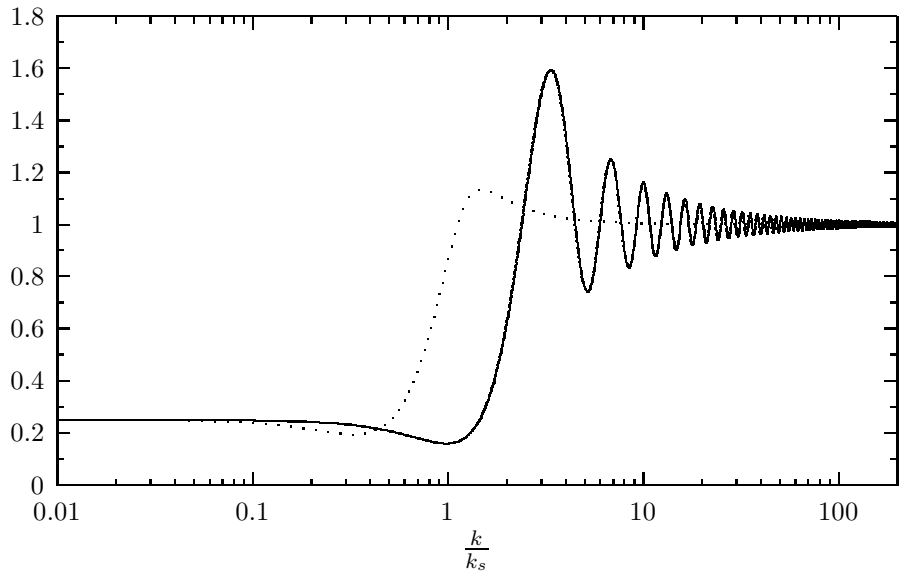} }
  \end{center}
  \caption[]{The quantities $\sigma^2_H(t_k)$ (dotted line) and
$k^3\Phi^2(k,t_k)\propto \delta^2_H(t_k)$ (solid line) are 
displayed for the same parameter value $p=0.5$. The overall 
normalization is arbitrary (and different) for each curve. 
It is obvious that the two quantities $\sigma^2_H(t_k)$ and
$\delta^2_H(t_k)$ have very different shapes in the vicinity 
of the characteristic scale $k_s$ where $\sigma^2_H(t_k)$ 
exhibits a spiky structure. Their respective maxima have 
different locations. Note the same asymptotic ratio between 
large and small values of $k$ (with $t_k \ll t_{eq}$).} 
\label{spectrum}
\end{figure}

We must distinguish three mass ranges: $M\ll M_{peak}$, $M\sim M_{peak}$ and 
$M \gg M_{peak}$. 
For large masses $M\gg M_{peak}$, $\beta(M)$ is unsignificant;
in this range we have a 
(nearly) flat plateau normalized to COBE.
For small masses, $M\ll M_{peak}$, we are on the 
higher plateau satisfying (\ref{beta2}) (see below). 
Note that on both plateaus,
corresponding to an effective spectral index $n=1$, 
PBHs of arbitrary mass are equally probable in view of (\ref{omega}).
The probability $\beta(M)$ that is obtained using (\ref{omega},\ref{sig})
displays a pronounced bump around the mass scale $M_{peak}\approx 0.6~M_s$,
see Fig.~2.  

The crucial property is the drop on smaller scales $M\ll M_{peak}$ which,
as we will show, is sufficient in order to satisfy the severe upper bounds set
on the contribution of evaporated PBHs to the diffuse $\gamma$-ray background.

Using the gravitational constraint (\ref{omega}), it is possible, for each 
value of the parameter $M_s$ to find the value of $p\geq p_{min}$
 which yields $\Omega_{PBH,0}\leq 0.3$. 
This is shown in Fig.~3.
Let us consider now the range of masses which have already evaporated
 by now and that contribute 
to the $\gamma$-ray background. Actually only a relatively small range
 $\Delta M$ with masses 
$2\times 10^{13}\ {\rm g}\leq M \leq 5\times 10^{14}\ {\rm g}$ contributes
significantly to the observed diffuse $\gamma$-ray background \cite{KL99}.
The fraction of energy density going into PBHs in this mass range obeys
\be
\beta_{-} \equiv \beta(M\ll M_{peak}) = \frac{\Omega_{PBH,0}
(M\ll M_{peak})}{\Omega_{PBH,0}(M_{peak})}~
\left ( \frac{M}{M_{peak}}\right)^{\frac{1}{2}}~\beta(M_{peak})~.
\ee
Taking $\Omega_{PBH,0}(M_{peak})=0.3$, this leads to the constraint
\be
\beta_{-} < 3.3 \times 10^{-8}~ \left
 ( \frac{M}{M_{peak}}\right)^{\frac{1}{2}}~\beta(M_{peak})~.\lb{betap}
\lb{beta2}
\ee
Note that we have used here the stronger constraint 
$\Omega_{PBH,0}(M\ll M_{peak}) < 10^{-8}$ for each scale $M\ll M_{peak}$
rather than the (integrated) constraint $\Omega_{PBH,0}(\Delta M) < 10^{-8}$. 

A shift in the location of the characteristic mass $M_s$
(and corresponding scale $k_s$) 
to higher masses (still much smaller than $10^{-7}\ {\rm M}_{\odot}$)
is allowed, but smaller values of $p$ will then be needed
in order to achieve $\Omega_{PBH,0}=0.3$. 
However, this in turn increases the ratio $\frac{\beta_-}{\beta_{peak}}$
which therefore gives an 
upper bound to $M_{peak}$ around $10^{21}\ {\rm g}$ as can be seen on Fig.~3. 
Hence, requiring that $\Omega_{PBH,0}=0.3$,
the $\gamma$-ray background constraint can be 
satisfied for a range of characteristic masses $M_s$
(with $M_s\approx 1.6~M_{peak}$) 
\be
10 M_*\lesssim M_s \lesssim 10^{21}\ {\rm g}~.\lb{Ms} 
\ee
For higher $M_{peak}$, the $\gamma$-ray background constraint
is no longer satisfied. 
An observational constraint on the allowed range of MACHO masses
would therefore 
give a strong constraint on our model as it must
correspond to the window (\ref{Ms}).

The bump for $M\simeq M_{peak}$ is well localized,
hence the PBH masses in this model indicate 
essentially the location of the characteristic scale $M_s$
in the primordial spectrum. 
Besides the various improvements we have used for the calculation
of PBH production, we note 
that the model presented here contains one singularity in the inflaton 
potential, in contrast to the more sophisticated version
with two bumps considered in \cite{INN94}.

It is interesting to compare these results with the approach
 in which PBHs of various masses $M$ 
are produced by near critical collapse \cite{NJ98} at one single time, 
the horizon crossing time $t_k$ corresponding to $M_H(t_k)=M_{peak}$, with
\be
M = K ( \delta - \delta_c )^{\gamma}~.\lb{cc}
\ee
One finds for $\gamma=0.35$, following \cite{Y98}, 
\be
\frac{d \Omega_{PBH}(M,t_{k_{peak}})}{d M}
\simeq 3.86~\frac{\beta(M_{peak})}{M_{peak}} 
          \left ( \frac{M}{M_{peak}} \right )^{2.86}~
\exp [ -1.35 \left (\frac{M}{M_{peak}} \right )^{2.86} ] ~.\lb{Y} 
\ee
This is turn yields 
\be
\Omega_{PBH,tot}(t_{k_{peak}}) = 0.8 ~\beta(M_{peak})~,\lb{omcc}
\ee
a result analogous to that obtained when PBHs of various mass
$M$ are produced at different times 
$t_k$ with $M=M_H(t_k)$; in the latter case one obtains 
$\Omega_{PBH,tot}(t_{k_{peak}}) = \beta(M_{peak})$. Hence, our constraint 
on $\Omega_{PBH,0}(M_{peak})$ using (\ref{omega}) 
is stronger than the requirement 
$\Omega_{PBH,tot}(t_0)\leq 0.3$ based on (\ref{omcc}).
Note that $M_{peak}$--$M_H$ in the notation of \cite{Y98}--
 is the Hubble mass corresponding 
to the peak in the quantity $\beta(M)$, not in the primordial 
spectrum itself, due to the 
distinction one has to make between the quantities $\sigma_H(t_k)$ 
and $\delta_H(t_k)$.
We stress also that the above mentioned approach does not take 
into account the properties 
of the feature in the primordial spectrum, in particular the 
width of the corresponding bump in 
$\beta$, besides the value $\beta(M_{peak})$ which serves as a 
kind of overall normalization of the 
PBH abundance. On the other hand, it does account in a consistent way
 for the production of 
PBHs also in the monotonically increasing part of $\beta(M)$
around $M_{peak}$, $M\lesssim M_{peak}$. 
Obviously, the more spiky the primordial spectrum, the better this approach is 
expected to be, and our model provides such a concrete spectrum 
based on an underlying 
inflationary dynamics. An accurate determination of the mass spectrum 
would therefore yield valuable 
information, not only on the underlying 
primordial perturbations spectrum but also on PBH formation itself. 

In conclusion, we have considered a power spectrum of inflationary origin in 
which PBHs constitute a significant part of CDM. 
In fact, it has already been speculated that 
the QCD phase transition could lead to PBHs of about one solar mass
\cite{KJ}, in accordance with microlensing observations made some
years ago. In our model PBHs are generated with much smaller mass.
Our analysis is based on an accurate calculation 
of the mass variance at the PBH formation 
time $t_k$ and of the corresponding $\beta(M)$.
Though it was found recently that a more accurate computation 
of the probability to produce PBHs 
actually leads to a decrease of this probability, 
this letter shows that one can still 
have a significant $\Omega_{PBH,0}$ by using primordial spectra 
in concrete models with a 
characteristic scale on very small scales, 
scales much smaller than those usually 
probed by CMB anisotropy or large scale structure formation.

\begin{figure}
  \begin{center} 
\mbox{ \input{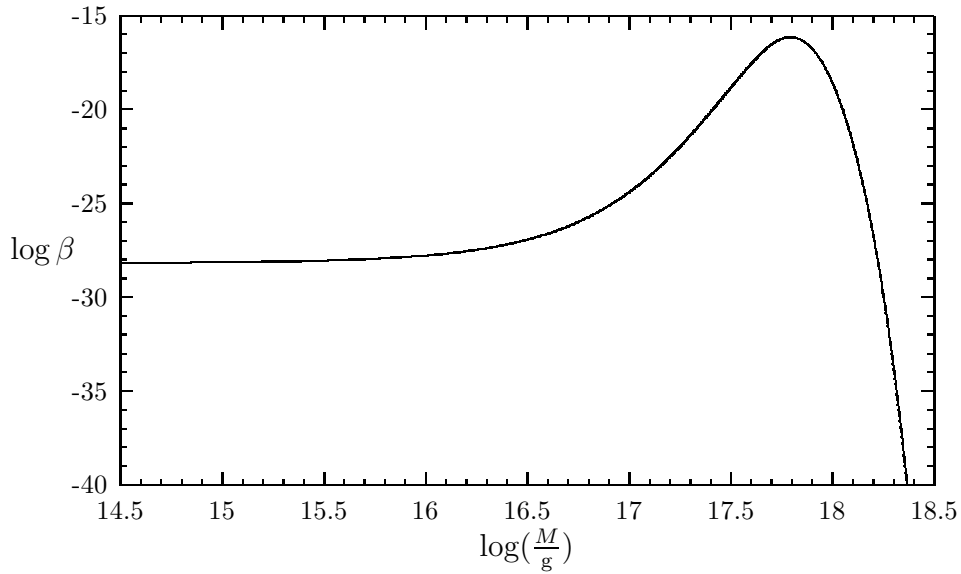} }
  \end{center}
  \caption[]{The quantity $\beta(M)$ is shown for our model 
containing a jump in the inflaton potential derivative for 
the parameters $p=7.913\times 10^{-4},~M_s=10^{18}{\rm g}$. 
As can be seen, $\beta(M)$ acquires a well localized bump 
in the vicinity of $M_s\equiv M(t_{k_s})$. Note that we have 
$M_s\approx 1.6~M_{peak}$ and $\beta_-\equiv \beta(M\ll M_{peak})
\approx 10^{-12}\times \beta(M_{peak})$, a value sufficient 
to avoid the severe constraint from the contribution to the 
$\gamma$-ray background of evaporated PBHs with $M\leq M_*$.}
\label{betafig}
\end{figure}

\begin{figure}
  \begin{center} 
\mbox{ \input{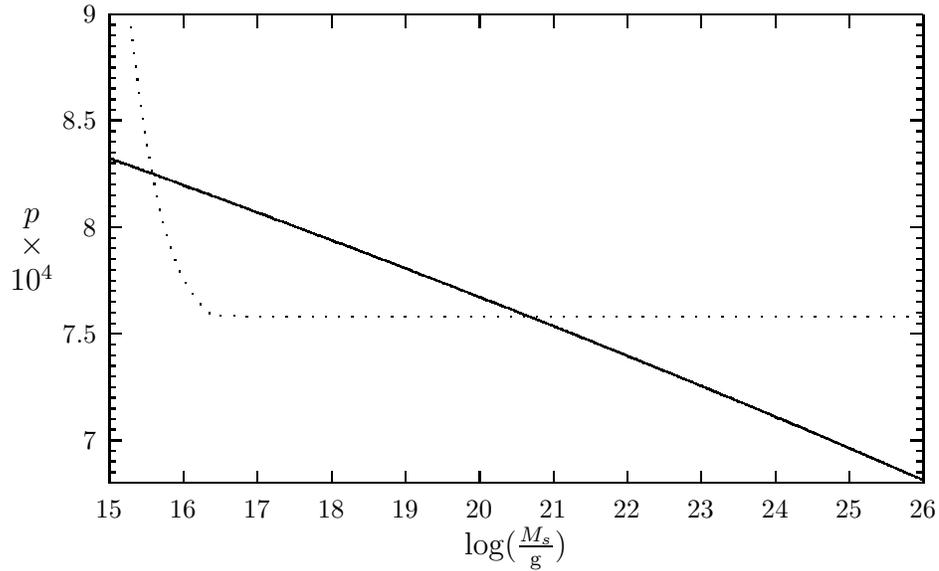} }
  \end{center}
  \caption[]{The allowed region in parameter space $(p, M_s) $ 
is shown. The solid line indicates the points for which 
$\Omega_{PBH,0}(M_{peak})=0.3$. Below the solid line, the 
gravitational constraint at $M_{peak}$ is violated and this 
region is therefore excluded. Below the dotted line, the 
$\gamma$-ray background constraint is violated. It is seen that 
for $10^{21}{\rm g}\lesssim M_s$, the allowed parameter values $p$ 
yield $\Omega_{PBH,0}(M_{peak})<0.3$ which becomes rapidly 
negligible with growing $M_s$. For the values of $p$ shown here, 
$M_s\approx 1.6 ~M_{peak}$.}
\label{constraint}
\end{figure}

\end{document}